\documentclass[sigconf,nonacm]{acmart}

\usepackage{tabularx}
\usepackage{multirow}

\usepackage{tcolorbox}
\usepackage{listings}
\usepackage{xcolor}
\usepackage{float}

\usepackage[ruled,vlined]{algorithm2e}

\usepackage{tikz}
\usetikzlibrary{shapes.geometric, arrows}
\tikzstyle{startstop} = [rectangle, rounded corners, minimum width=3cm, minimum height=1cm,text centered, draw=black, fill=red!30]
\tikzstyle{process} = [rectangle, minimum width=3cm, minimum height=1cm, text centered, draw=black, fill=orange!30]
\tikzstyle{decision} = [diamond, minimum width=3cm, minimum height=1cm, text centered, draw=black, fill=green!30]
\tikzstyle{arrow} = [thick,->,>=stealth]

\begin{document}

\title[SLA Management in Reconfigurable Multi-Agent RAG]{SLA Management in Reconfigurable Multi-Agent RAG: A Systems Approach to Question Answering}

\author{Michael Iannelli}
\affiliation{%
  \institution{Yext, Inc.}
  \city{New York}
  \state{New York}
  \country{USA}}
\email{miannelli@yext.com}

\author{Sneha Kuchipudi}
\affiliation{%
  \institution{Yext, Inc.}
  \city{New York}
  \state{New York}
  \country{USA}}
\email{skuchipudi@yext.com}

\author{Vera Dvorak}
\affiliation{%
  \institution{Yext, Inc.}
  \city{New York}
  \state{New York}
  \country{USA}}
\email{vdvorakova@yext.com}

\renewcommand{\shortauthors}{Iannelli et al.}

\begin{abstract}
Retrieval Augmented Generation (RAG) enables Large Language Models (LLMs) to generalize to new information by decoupling reasoning capabilities from static knowledge bases. Traditional RAG enhancements have explored vertical scaling—assigning subtasks to specialized modules—and horizontal scaling—replicating tasks across multiple agents—to improve performance. However, real-world applications impose diverse Service Level Agreements (SLAs) and Quality of Service (QoS) requirements, involving trade-offs among objectives such as reducing cost, ensuring answer quality, and adhering to specific operational constraints.

In this work, we present a systems-oriented approach to multi-agent RAG tailored for real-world Question Answering (QA) applications. By integrating task-specific non-functional requirements — such as answer quality, cost, and latency — into the system, we enable dynamic reconfiguration to meet diverse SLAs. Our method maps these Service Level Objectives (SLOs) to system-level parameters, allowing the generation of optimal results within specified resource constraints.

We conduct a case study in the QA domain, demonstrating how dynamic re-orchestration of a multi-agent RAG system can effectively manage the trade-off between answer quality and cost. By adjusting the system based on query intent and operational conditions, we systematically balance performance and resource utilization. This approach allows the system to meet SLOs for various query types, showcasing its practicality for real-world applications.
\end{abstract}

\ccsdesc[500]{Information systems~Information retrieval}
\ccsdesc[500]{Applied computing~Service-oriented architectures}
\ccsdesc[500]{Computing methodologies~Distributed artificial intelligence}

\keywords{Retrieval Augmented Generation, Multi-Agent Systems, Service Level Agreement}

\maketitle
\section{Introduction}
Question Answering (QA) has been a pivotal area of research since the 1960s, aiming to provide precise answers to natural language questions. Unlike traditional search engines that return a list of hyperlinks to relevant pages, QA systems deliver direct answers to user queries \cite{zhu-qa}.

The advent of generative Large Language Models (LLMs) has led search engines to integrate abstractive QA into their offerings—systems that generate answers which may not exist verbatim in source documents \cite{AIPowered_2024}. Additionally, new platforms specializing in abstractive QA, such as Perplexity \cite{perplexity-rag} and ChatGPT Search \cite{chatgpt-search}, have emerged.

However, LLM-based abstractive QA exhibits notable limitations, including a lack of generalization to unseen data and a propensity for hallucination, i.e. the generation of false assertions \cite{survey_hallucination}.  Retrieval Augmented Generation (RAG) \cite{gao2024ragllm} addresses these shortcomings by combining LLMs with external knowledge sources, enabling the LLM to function as a reasoning layer rather than a static repository of facts. RAG retrieves context from external sources such as search engines, traditional databases, vector databases \cite{jing2024llm-vector}, or knowledge graphs \cite{knowledge-injection, xu-rag-customer}, and utilizes the LLM for reasoning and response generation.

Enhancements to RAG-based QA have primarily focused on decomposing the task into smaller subtasks and distributing responsibilities among specialized components. While these techniques have improved answer quality, they often introduce additional costs in terms of computation or increased response times.
    
Real-world QA systems must navigate trade-offs between answer quality and the costs associated with providing those answers. Moreover, they must be dynamically reconfigurable to handle diverse Service Level Agreements (SLAs) and Quality of Service (QoS) requirements arising from various query intents and problem domains. For instance:

\begin{itemize}
    \item \textbf{E-commerce Search:} Highly sensitive to response times, where delays can significantly impact user experience and sales conversions \cite{vanvessum_amazon_page_speed_study, achary_latency_impact_ecommerce}.
    \item \textbf{Customer Support:} Requires not just correct answers but stylistically compliant ones that adhere to the brand-voice guidelines of the domain or company \cite{brandvoice}.
    \item \textbf{Legal and Healthcare Queries:} Demand high answer quality due to the critical nature of the information, with less emphasis on cost or latency constraints \cite{legalrag}.
\end{itemize}

Furthermore, because modern cloud providers (e.g., SaaS and PaaS platforms) host these varied workloads side by side on shared infrastructure, the system’s overall utility is increased: by optimally provisioning and multiplexing resources across collocated services with diverse SLA requirements, we achieve higher utilization and more efficient service delivery.  
  
In addition to these domain-specific requirements, real-world RAG systems operate under constantly changing and sometimes adverse conditions. Factors such as network congestion, computational capacity constraints, and security concerns necessitate adaptive strategies. For example, during periods of high network traffic, minimizing external API calls and leveraging local LLM instances can help maintain answer quality while adhering to operational constraints.

By incorporating knowledge about the task domain, QoS requirements, and the operational environment, we can quantify and manage the trade-offs between cost, answer quality, and other non-functional requirements. This enables the design of QA systems that are both efficient and effective across a range of real-world scenarios.

\subsection{Our Contributions}
We present a novel system-theoretic framework for SLA management in reconfigurable multi-agent RAG systems, specifically tailored for question-answering applications. Our main contributions are as follows:
\begin{itemize}
    \item \textbf{Dynamically Reconfigurable Horizontal Scaling Framework}: We propose a method for horizontally scaling multi-agent RAG systems by replicating agents with diverse configurations. This approach systematically improves performance by enabling the system to dynamically adjust resource allocation based on the specific requirements of each query.
    \item \textbf{Implementation and Experimental Validation}: We detail the architecture and implementation of our Abstractive Question Answering Intent Handler. Additionally, we present experimental results that demonstrate the effectiveness of our approach, including comparisons of strategies employed by different agent architectures.
    \item \textbf{Novel Metrics and Dataset Creation}: We introduce novel metrics and a dataset creation process to address stylistic adherence requirements, which are critical for many industry users.
\end{itemize}

By addressing the challenges of SLA management in multi-agent RAG systems, our work contributes to the development of QA systems that are adaptable, efficient, and capable of delivering high-quality answers under diverse operational conditions.

\subsection{Related Work}

Scaling Retrieval-Augmented Generation (RAG) systems has been an active area of research, with most approaches focusing on vertical scaling—decomposing tasks into subtasks assigned to specialized components. Recent studies incorporate Large Language Models (LLMs) at the planning stages to mitigate issues like irrelevant retrieval and focus drift. For instance, PlanRAG uses an LLM to generate retrieval plans, enhancing context relevance \cite{planrag}. Similarly, Retrieve-Plan-Generate (RPG) separates planning and answering stages to improve long-form question answering \cite{rpg2024}. AutoGPT+P introduces planning loops for step-by-step plan generation in dynamic environments \cite{autogpt+p}.

Horizontal scaling through multi-agent systems has also been explored, where multiple agents are synchronously assigned to the same question. LLM-Debate employs multiple LLM agents debating to reach better answers \cite{llm-debate}. Chain-of-Thought Self-Consistency (CoT-SC) generates multiple reasoning paths and selects the most consistent answer \cite{cot-sc}. Li et al. demonstrate that increasing the number of agents with simple voting improves performance \cite{moreagents}.

Temporal scaling techniques involve sequential interactions with LLMs to enhance question-answering performance on complex queries. Shao et al. use iterative querying to refine answers for complex information needs \cite{iterative-retrieval}. Query refinement methods, such as RQ-RAG involve rewriting, decomposing, and disambiguating complex queries have also been shown to improve answer quality \cite{rq-rag}.

While these aforementioned methods enhance answer quality, they often overlook dynamic adaptation to operational constraints critical in real-world applications with strict Service Level Agreements (SLAs).

Gao et al. introduce a modular RAG framework that encompasses horizontal, vertical, and temporal scaling techniques \cite{modular-rag}. Zhang et al. propose an agentic information retrieval framework capable of handling complex queries by introducing IR agents with greater levels of autonomy \cite{agentic-ir}. However, these works lack experimental results on adapting systems to real-world environments.

Arbitration mechanisms have been proposed to improve decision-making in multi-agent systems. Some studies suggest using multiple LLMs to select the best response \cite{PoLL}. These methods improve answer quality but may not address trade-offs such as cost, latency, and security.

Evaluation frameworks like RAGAS assess RAG performance but may omit industry-relevant metrics such as adherence to stylistic constraints \cite{ragas}. In practical applications, an answer must be not only correct but also stylistically appropriate.

Our work differs by focusing on the dynamic reconfiguration of multi-agent RAG systems to meet diverse SLAs in real-world settings. We integrate task-specific non-functional requirements into system parameters, addressing trade-offs between answer quality, cost, latency, and adherence to stylistic guidelines. This practical framework aims to deploy question-answering systems in industrial contexts with stringent SLA demands.

\section{SLA Composition in Multi-Agent QA Systems}
We adopt a compositional perspective on Service Level Agreements (SLAs), decomposing a complex SLA into multiple interacting Quality-of-Service (QoS) attributes—such as response latency, factual accuracy, stylistic adherence, and cost—and then combining them into an overall service commitment. Under this framework, the composite SLA for a QA system is defined as

\[
\mathrm{SLA}_{\mathrm{composite}} = f(q_1, q_2, \dots, q_k),
\]

where each \(q_i\) denotes a QoS attribute (e.g., latency, answer quality, stylistic compliance) and the function \(f\) captures how these attributes interact.

In our reconfigurable multi-agent RAG architecture, every module—from Intent Detection and Planning to individual QA agents and the Arbitration component—contributes toward one or more of these QoS attributes. For example, the Planning Module dynamically allocates both the number and configuration of RAG agents based on the incoming query’s requirements and current operational conditions; this allocation explicitly considers the relevant SLA sub-attributes to ensure the final answer meets its targets for speed, accuracy, and style.

A continuous monitoring subsystem then audits each module’s compliance with its assigned sub-SLAs in real time. Whenever a deviation from the target QoS levels is detected—whether in response time, factual correctness, or stylistic conformity—adaptive reconfiguration is triggered. Such adjustments might include tuning the arbitration thresholds, shifting workload between cloud and local LLM instances, or reallocating compute resources. This feedback loop ensures that the composite SLA is upheld by continuously reconciling the performance of individual components.

By embedding this compositional SLA framework into our multi-agent QA system, we achieve a systematic balance among answer quality, latency, and cost—yielding a robust, adaptable service that satisfies diverse and stringent SLA requirements in real-world deployments.

\section{Architecture}
\begin{figure}[htbp]
\centerline{\includegraphics[width=.85\linewidth]{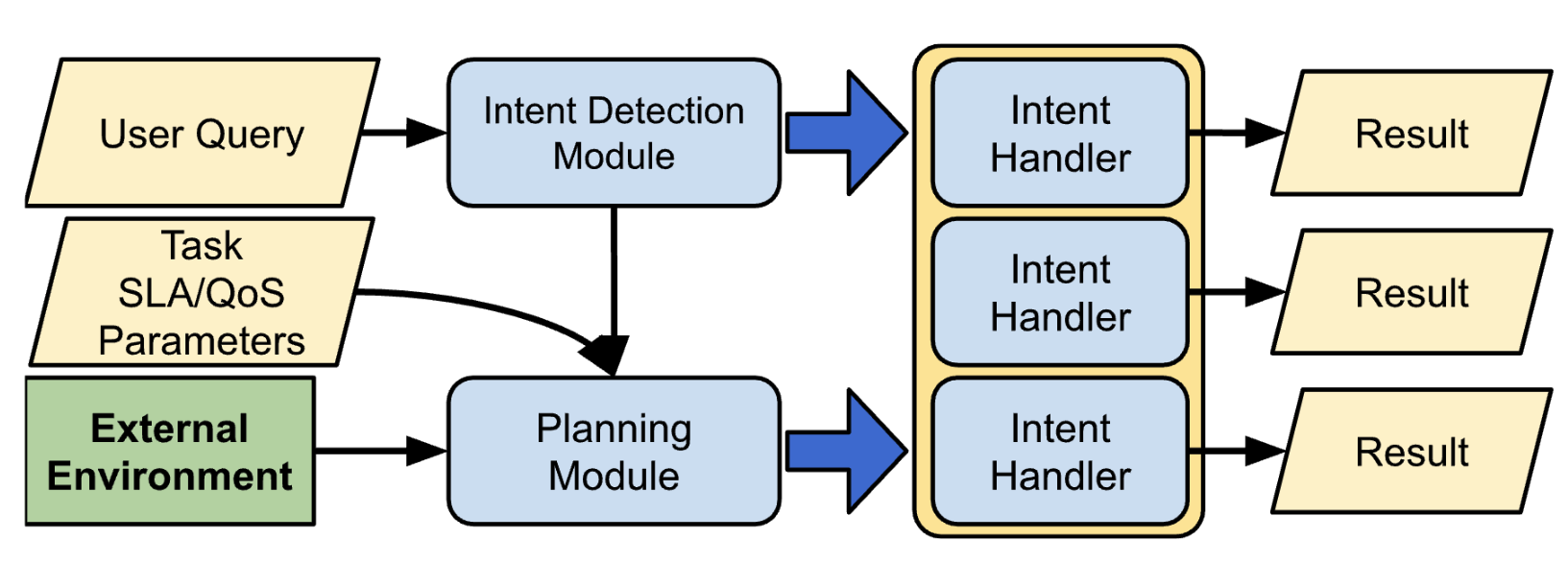}}
\caption[A bird's eye view of the architecture of the AbstractiveQA System]{A bird's eye view of the architecture of the AbstractiveQA System}
\label{high_level_arch}
\end{figure}

Our Question Answering System is designed to dynamically reconfigure its components to meet Service Level Agreements (SLAs). The system consists of three primary modules: an Intent Detection Module, a Planning Module, and multiple Intent Handlers. A high-level overview of the architecture is shown in Figure~\ref{high_level_arch}.

\subsection{System Inputs and Outputs}
\subsubsection{User Query}: The system accepts arbitrary input queries from users, such as: ``How do I reset my phone?", ``How to reset phone", or ``Burger joints near me".

These queries vary significantly in intent and complexity, necessitating a robust mechanism for accurate intent recognition and appropriate handling.

\subsubsection{Task SLA/QoS Parameters}: Specific requirements such as:
\begin{itemize}
    \item Required precision and recall for the task
    \item Maximum acceptable hallucination rate and incongruent response rate (to ensure adherence to stylistic constraints)
    \item Budget and Latency constraints for different intent types
\end{itemize}

\subsubsection{Operational Environment}
Current conditions affecting the system, such as:
\begin{itemize}
    \item Availability of network resources
    \item Status of external APIs and services
    \item Available models and data sources
\end{itemize}

\subsubsection{Results}: The QA system's outputs are \textbf{Intent-Dependent Answers} which consist of diverse responses tailored to the user's intent and complying with specified SLAs.

\subsection{Intent Detection Module}
The Intent Detection Module processes the user query to classify it into predefined intent categories, focusing on determining the \textit{what} of the input query\footnote{The \textit{what} refers to the type or kind of query being made, as opposed to the \textit{why}, which pertains to the motivation behind the user's question.}, i.e., what kind of query it is:
\begin{itemize}
    \item \textbf{Directly Answerable Question:} Queries seeking specific information that can be directly addressed.
    \item \textbf{Request for Summarization:} Queries asking to condense or summarize information from a source.
    \item \textbf{Non-Question Statements:} Inputs that are not interrogative but may require action.
    \item \textbf{Request for a List:} Queries seeking a collection of items (e.g., nearby restaurants serving hamburgers).
    \item \textbf{Sales Inquiry:} Queries related to purchasing or product information.
\end{itemize}

\subsubsection{Implementation Options}
Implementation options for this module include various techniques, each with trade-offs in terms of computational cost, extensibility, and accuracy\footnote{In this context, \textit{accuracy} is used in a qualitative sense to measure the quality of outcomes, rather than the quantitative definition commonly used in machine learning.}:
\begin{itemize}
    \item \textbf{Heuristic Methods:} Rule-based systems using predefined patterns and keywords for quick intent classification. While low in computational overhead, they may lack flexibility and scalability.
    \item \textbf{Machine Learning Classifiers:} Lightweight models trained on labeled datasets to predict intents, offering a balance between performance and resource utilization.
    \item \textbf{Generative Large Language Models (LLMs):} Zero-shot or few-shot classifiers leveraging pre-trained LLMs to understand and classify intents without extensive domain-specific training data. Although highly extensible and accurate, they incur higher computational costs.
\end{itemize}

\subsection{Planning Module}
The Planning Module dynamically configures the resources of the Intent Handlers to meet the system's SLA requirements for each specific intent in the current environment. Its core functions include:
\begin{itemize}
    \item Resource Allocation: Determining the optimal number of Retrieval Augmented Generation (RAG) agents to deploy, balancing response quality against computational cost.
    \item Backend Data Source Selection: Choosing appropriate data repositories (e.g., cached datasets, real-time databases, knowledge graphs) based on required information freshness and relevance.
    \item Arbitration Mechanism Configuration: Selecting and parameterizing arbitration algorithms to aggregate and evaluate responses from multiple agents effectively.
\end{itemize}

For example, to service a query requiring high answer quality, the Planning Module may:
\begin{itemize}
\item Increase Agent Replication: Deploy more RAG agents to generate diverse candidate answers.
\item Employ Appropriate Data Sources: Access multiple data repositories to enrich the context and determine which agents should have access to which sources e.g. an agent using a third-party LLM would not get access to sensitive documents.
\item Select what threshold to use in a voting-based arbitration mechanism e.g. only return an answer when 70\% of the agents think it should be answered.
\end{itemize}

By adjusting these configurations in real-time, the Planning Module ensures compliance with SLA parameters, optimizing the system's performance in terms of answer quality and resource utilization.

\begin{figure}[htbp]
\centerline{\includegraphics[width=0.85\linewidth]{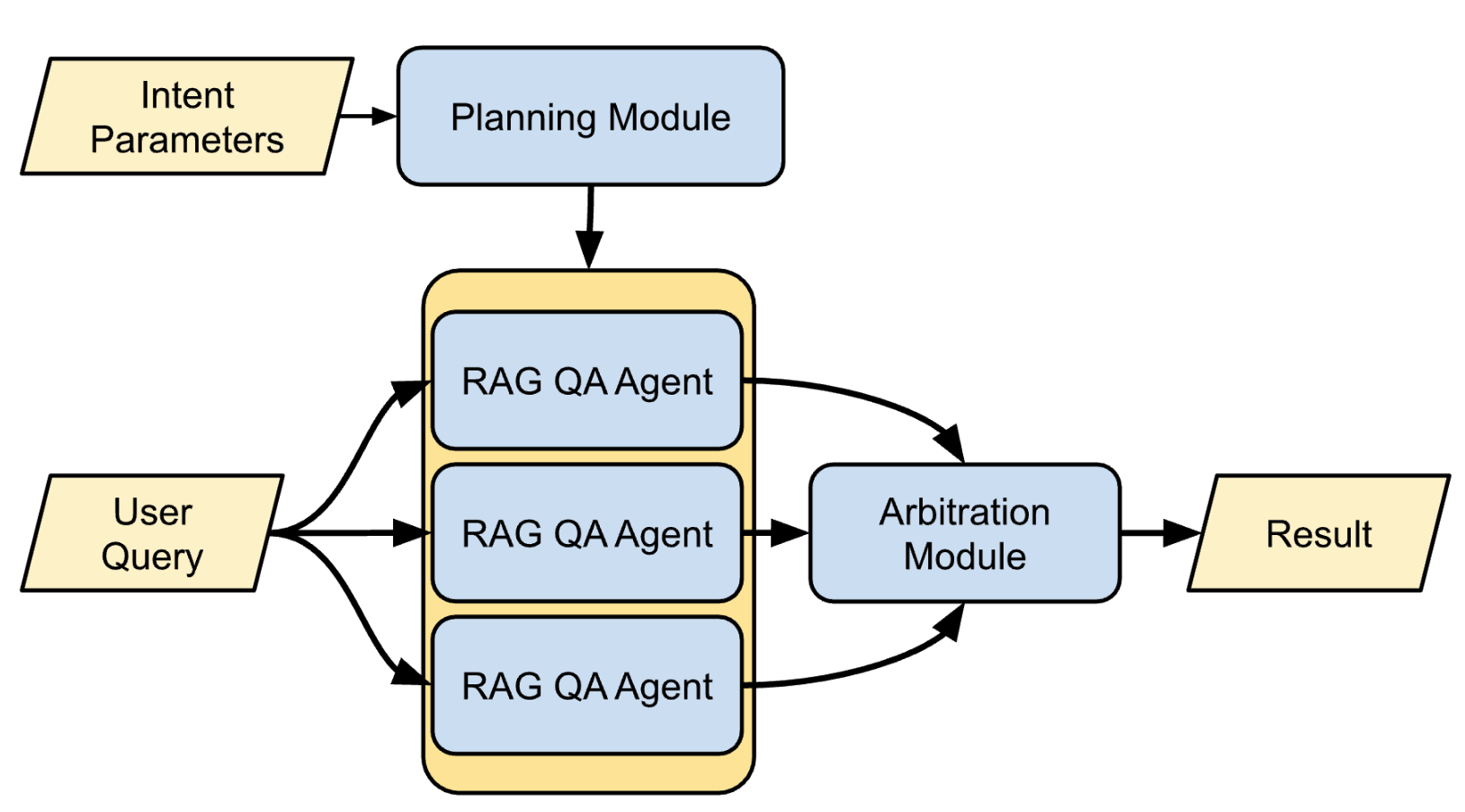}}
\caption{Possible Instantiation of Abstractive Question Answer Intent Handler}
\label{intent_handler_arch}
\end{figure}

\subsection{Intent Handlers}

The Intent Handler is a reconfigurable module that executes the necessary processes to address the user's query based on the classified intent. By dynamically adjusting its internal workflows and resource utilization in response to directives from the Planning Module, the Intent Handler can:

\begin{itemize}
    \item \textbf{Manage Computational Costs}: Allocate resources judiciously to stay within budgetary constraints without significantly compromising answer quality.
    \item \textbf{Enhance Answer Quality}: Allocate additional resources, such as more RAG agents or advanced arbitration mechanisms, for queries where accuracy is paramount.
    \item \textbf{Ensure Compliance with Non-Functional Requirements}: Adjust processing to adhere to stylistic guidelines, security policies, and other non-functional requirements specified in the SLAs.
\end{itemize}

To achieve these objectives, the Intent Handler orchestrates a network of logical entities and services, which may include:

\begin{itemize}
    \item \textbf{Backend Data Access Interfaces}: APIs and connectors to interact with various data sources.
    \item \textbf{RAG Agents}: Multiple instances of Retrieval Augmented Generation agents, each potentially configured with different retrieval strategies, preprocessing pipelines, and LLMs to generate diverse candidate responses.
    \item \textbf{Arbitration Algorithms}: Mechanisms to evaluate and select the optimal response from the candidate answers. This may involve:
    \begin{itemize}
        \item \textbf{Voting Systems}: Methods like majority voting or weighted voting based on agent confidence scores.
        \item \textbf{Ranking Models}: Machine learning models trained to rank candidate responses based on relevance, accuracy, and other quality metrics.
    \end{itemize}
    \item \textbf{Ensemble Techniques}: Combining outputs from multiple agents to produce a more robust final answer.
    \item \textbf{Response Synthesis}: Constructing the final response, ensuring coherence and adherence to stylistic guidelines.
\end{itemize}

A diagram showing a potential instantiation of a multi-agent Abstractive QA Intent Handler is presented in Figure~\ref{intent_handler_arch}.

\section{Benchmark Dataset Creation and Metrics}

\subsection{Dataset Creation}
We sampled approximately 5,000 actual user queries spanning over 200 different businesses across various industries (e.g. retail, finance, etc.) Each query was classified into one of two intent categories: \textbf{Directly Answerable Questions} or \textbf{All Other Intents}.

In this context, a query is considered directly answerable if it can be addressed succinctly (e.g., with a yes/no or factual response), such as inquiries about a product, service, or specific company information. Non-directly answerable queries include broad topics, technical issues, or those better answered with a list or document link.

Each query was annotated by two independent annotators, followed by a conflict resolution process when necessary. The resulting dataset served two purposes:
\begin{itemize}
\item Evaluation of the Intent Classifier (which is outside the scope of this research).
\item Creation of an annotated dataset for queries and answers.
\end{itemize}

We found that directly answerable queries constituted approximately 12\% of the original dataset, totaling around 600 queries. Each of these queries was input into five separately configured RAG-QA Agents, resulting in approximately 3,000 (query, context, answer) triples. Each triple was annotated by two independent annotators, with a conflict resolution step to ensure consistency.

Each result was assigned three attributes reflecting: (i) the availability of an answer in context, (ii) the provision of an answer, and (iii) the quality of the answer. Detailed definitions for these attributes are provided in Table~\ref{tab:qa_annotation}.

{
  \setlength{\tabcolsep}{8pt}      
  \renewcommand{\arraystretch}{1.2} 

  \begin{table*}[h]
  \centering
  \begin{tabularx}{\textwidth}{|>{\raggedright\arraybackslash}p{2.5cm}|>{\raggedright\arraybackslash}p{3cm}|X|}
  \hline
  \textbf{Category} & \textbf{Option} & \textbf{Description} \\ \hline
  \multirow{2}{2.5cm}{\raggedright Context Availability} 
      & Answer Exists in Context 
      & At least one agent’s retrieval process returned a result from which a correct direct answer could be derived. \\ \cline{2-3}
      & Answer Does Not Exist in Context 
      & None of the agents’ retrieval processes returned a result from which a correct direct answer could be derived. \\ \hline
  \multirow{2}{2.5cm}{\raggedright Answer Provision} 
      & Answer Provided 
      & An answer was returned (regardless of its correctness). \\ \cline{2-3}
      & No Answer Provided 
      & A negative, or null, answer was returned. \\ \hline
  \multirow{3}{2.5cm}{\raggedright Answer Quality} 
      & Correct Answer 
      & The returned answer was correct and aligned with the task’s non-functional requirements (e.g., adhered to the specified style in the prompt). \\ \cline{2-3}
      & Hallucination 
      & The returned answer was not grounded in the provided context. \\ \cline{2-3}
      & Incongruent Response 
      & The returned answer did not adhere to the task’s non-functional requirements. \\ \hline
  \end{tabularx}
  \caption{Summary of Answer Annotation Attributes}
  \label{tab:qa_annotation}
  \end{table*}
}

\subsection{Metrics}
\subsubsection{Answer Quality Metrics}
\label{sub:aq_metrics}

We evaluated the performance of the agents using the following metrics:

\begin{itemize}
    \item \textbf{Precision}: The number of correct answers divided by the total number of answers provided.
    \item \textbf{Recall}: The number of correct answers divided by the number of times the answer exists in the global context (i.e., the union of all contexts from all agents).\footnote{The use of the global context to compute precision and recall introduces limitations to our dataset, as the exact upper bound on the number of answers in any context is unknown and could not be determined within our time constraints. However, we found that the difference in the number of answers sourced from the global context between the four-agent and five-agent systems was minimal (one to two instances, depending on configuration). This suggests that the overall impact is likely negligible with the retrieval mechanisms used.}
    \item \textbf{F1 Score}: The harmonic mean of precision and recall:
    \[
    F_1 = 2 \cdot \frac{\text{Precision} \cdot \text{Recall}}{\text{Precision} + \text{Recall}}
    \]
    \item \textbf{Hallucination Rate}: The number of hallucinations divided by the total number of answers provided, where a hallucination is any answer not grounded in the context.\footnote{The Hallucination Rate is analogous to the False Positive Rate, with the key difference being that false positives in this context include answers not grounded in the provided context (extrinsic hallucinations), contradicted by the provided context (intrinsic hallucinations), or incongruent responses. Further details on categories of hallucination can be found in \cite{survey_hallucination}.}
    \item \textbf{Incongruent Response Rate}: The number of incongruent responses divided by the total number of responses.
\end{itemize}

\paragraph{Precision, Recall, and F1:} In traditional information retrieval, precision is defined as the proportion of retrieved documents that are relevant, and recall as the proportion of relevant documents that are retrieved \cite{Manning2008}. In our question answering domain, we adapted these definitions to suit our specific context.

\paragraph{Incongruent Responses:} For many users of this system, particularly brands, it is insufficient to simply return a correct answer; the answer must also adhere to the non-functional and stylistic requirements specified in the prompt. This includes guidelines such as not breaking the fourth wall or referencing the context directly. Ensuring compliance with these requirements is essential for maintaining brand voice and user experience.

\subsubsection{System Cost}
Our model incorporates a fixed overhead cost per question, the cumulative cost of the diverse QA agents deployed for the intent, and the associated arbitration cost. Mathematically, this is expressed as:

\[
C_{\text{sys}} = C_{\text{overhead}} + \sum_{i=1}^{N_{\text{intent}}} C_{\text{agent}, i} + C_{\text{arbitration}}(N_{\text{intent}})
\]

Where:
\begin{itemize}
    \item \(C_{\text{sys}}\): Total system cost of answering one question.
    \item \(C_{\text{overhead}}\): Fixed system overhead cost per question.
    \item \(C_{\text{agent}, i}\): Cost of the \(i\)-th QA agent.
    \item \(N_{\text{intent}}\): Number of QA agents deployed for the intent.
    \item \(C_{\text{arbitration}}(N_{\text{intent}})\): Arbitration cost for the \(N_{\text{intent}}\) agents.
\end{itemize}

\subsubsection{System Latency}
The system latency for our implementation is defined as the sum of the fixed overhead latency, the maximum latency among the QA agents configured for a given intent (since agent operations are asynchronous and parallel), and the arbitration latency for those agents. Mathematically, this is expressed as:

\[
L_{\text{sys}} = L_{\text{overhead}} + \max_{1 \leq i \leq N_{\text{intent}}} L_{\text{QA}, i} + L_{\text{arbitration}}(N_{\text{intent}})
\]

Where:
\begin{itemize}
    \item \(L_{\text{sys}}\): Total system latency.
    \item \(L_{\text{overhead}}\): Fixed overhead latency.
    \item \(L_{\text{QA}, i}\): Latency of the \(i\)-th QA agent.
    \item \(\max_{1 \leq i \leq N_{\text{intent}}} L_{\text{QA}, i}\): Maximum latency among all \(N_{\text{intent}}\) QA agents.
    \item \(L_{\text{arbitration}}(N_{\text{intent}})\): Arbitration latency for the \(N_{\text{intent}}\) agents.
\end{itemize}

\section{Agent Implementation}
In this section, we describe the experiments conducted to evaluate our reconfigurable multi-agent RAG system for question answering. We detail the agent architecture, single-agent experiments, and the arbitration algorithm, including its implementation and ablation studies.

\begin{figure}[htbp]

\centerline{\includegraphics[width=0.75\linewidth]{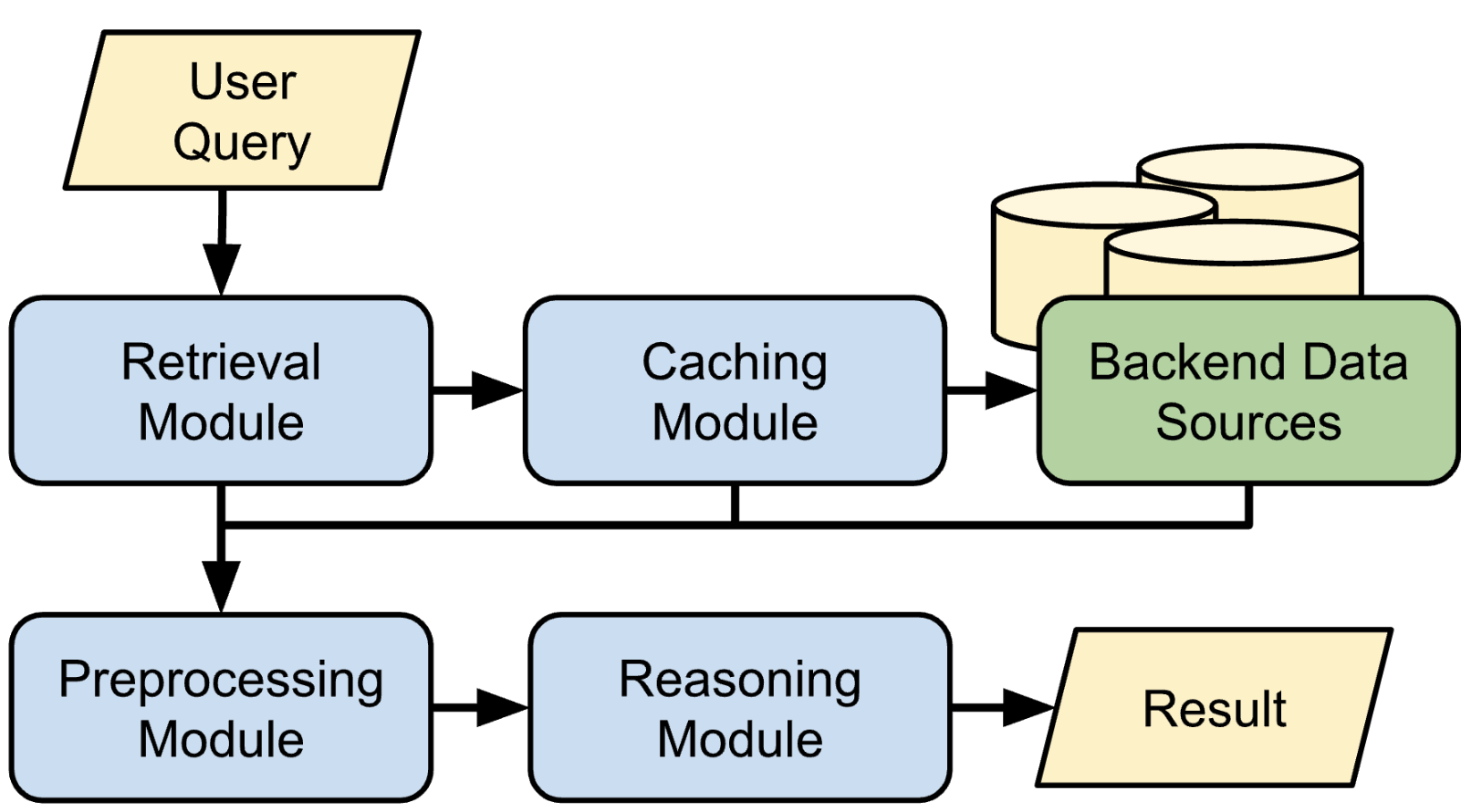}}
\caption{The RAG QA Agent Architecture we use in our experiments}
\label{agent_arch}
\end{figure}

\subsection{Agent Architecture}
The architecture of an individual QA agent used in our experiments, shown in Figure \ref{agent_arch}, is designed to be flexible and configurable based on the output of the Planning Module. The specific parameterization of each agent depends on factors such as which backend data sources are provisioned and caching strategies. While the implementation details of the Planning Module are outside the scope of this research, we provide an overview of the QA agent's architecture and workflow.

\begin{itemize}
\item Query Input: The user query is input into the Retrieval Module.  Only queries classified as directly answerable are processed by the QA agent.
\item Retrieval Module: Accesses backend data sources using the input query. Utilizes a federated search approach, querying multiple backend data sources asynchronously. Supports vertical search\footnote{Vertical search refers to specialized searches within predefined categories or domains. For instance, a healthcare-related query for cardiology may return results segmented into verticals such as cardiologists and hospitals specializing in cardiology.} by returning data categorized by entity type, configured based on domain requirements. Verticals are ranked in order of relevance.

\item Post-Retrieval Preprocessing: A preprocessing module prepares the context documents for input into the Reasoning Module. Prunes and organizes retrieved documents to comply with the LLM's context window size limitations.
\item Reasoning Module:  Powered by GPT-4 \cite{gpt4techreport}.  Generates answers based on the user query and preprocessed context.  Details of the prompt used are discussed in Section \ref{sub:prompt_engineering}.
\end{itemize}

\begin{figure*}[htbp]
\centerline{\includegraphics[width=\linewidth]{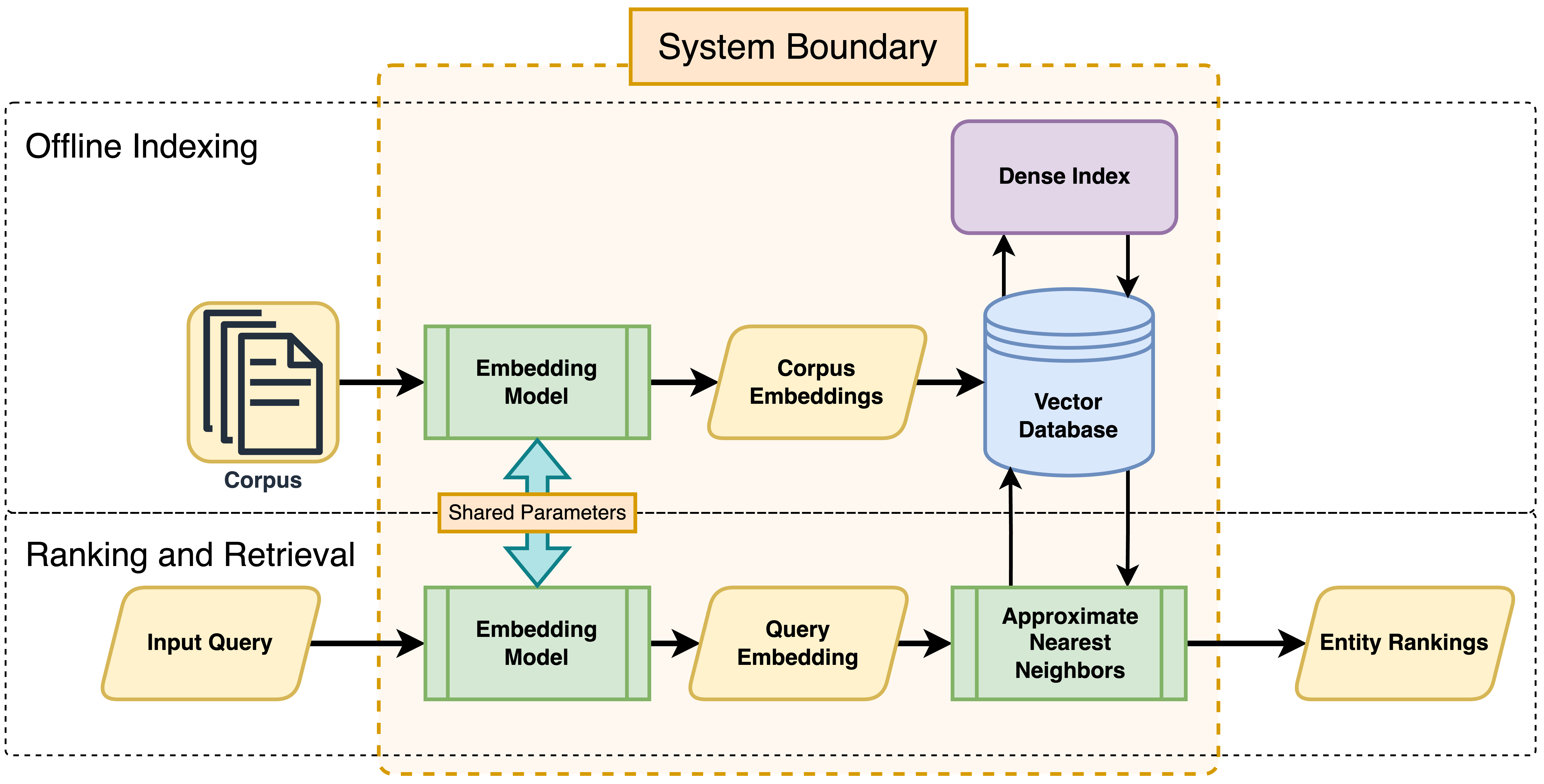}}
\caption{Architecture of the Vector Search Process we use in our experiments}
\label{vector_search}
\end{figure*}

\subsection{Retrieval Module Details}
The Retrieval Module, shown in Figure \ref{vector_search}, pulls data from various backend sources, including vector databases, lexical search engines, and graph data stores, depending on the domain. For vector search, we employ the following components:

\begin{itemize}
\item Embedding Model: Documents are embedded using a transformer model based on MPNet \cite{mpnet}.  Embeddings capture semantic information for effective similarity comparisons.
\item Vector Database: Indexed using Hierarchical Navigable Small World (HNSW) graphs \cite{hnsw} for efficient approximate nearest neighbor search.
At query time, the same embedding model is used to encode the user query. The resulting query vector $\mathbf{q}$ is used to retrieve documents whose vectors $\mathbf{d}$ are similar, based on cosine similarity:
$$
similarity(\mathbf{q}, \mathbf{d}) = \frac{\mathbf{q} \cdot \mathbf{d}}{\|\mathbf{q}\| \|\mathbf{d}\|}
$$
Documents with higher cosine similarity scores are considered more relevant to the query.
\end{itemize}

\subsection{Prompt Engineering}
\label{sub:prompt_engineering}
After re-ranking and/or thresholding, we construct the prompt for the LLM. The preprocessed list of context documents and the original user query are interpolated into the prompt at appropriate places. An example prompt is shown in Figure \ref{fig:prompt-example}.

\section{Experiments}

We conducted experiments on individual QA agents by varying the post-retrieval preprocessing logic used to prepare the input prompts for the GPT-4-powered Reasoning Module. Given the LLM's context window limitation of 8,000 tokens, we needed to prune the retrieved context appropriately. We developed several pruning methods aiming to balance content relevance with the token limit. For each agent, we measured Answer Quality using the metrics described in Section \ref{sub:aq_metrics}.

\begin{figure}[H]
\centering
\begin{tcolorbox}[
    colback=blue!5!white, 
    colframe=blue!75!black, 
    width=\linewidth, 
    title=Prompt Example,
    boxsep=2pt,    
    left=2pt, 
    right=2pt, 
    top=2pt, 
    bottom=2pt
]
\begin{lstlisting}[basicstyle=\linespread{1.1}\selectfont\ttfamily\small, aboveskip=0pt, belowskip=0pt]
   Context: {context_entities}.
   Use only the provided context to answer the 
   question: `{query}' to the best of your 
   ability and in a few sentences or less.
   
   If there is more than one answer, summarize 
   the options. Provide the `uid' values of the 
   object you used to inform the answer (do NOT 
   use the `id' value). Return the answer in a 
   json dictionary format 
    [Example: 
        {
            `answer': `This is the answer', 
            `uid_list': [12345, 98342]
        }]
   Do not refer to the context in your answer.
\end{lstlisting}
\end{tcolorbox}
\caption{An example of the constructed prompt used by the QA Agent's Reasoning Module.}
\label{fig:prompt-example}
\end{figure}

For the multi-agent experiments, we implemented an arbitration algorithm based on majority voting and relevance ranking. We varied the size of the agent ensemble dedicated to servicing the QA queries by adjusting the number of QA agents—each employing diverse post-retrieval preprocessing strategies—included in the ensemble. The final answer was selected based on the arbitration algorithm. Similar to the single-agent experiments, we measured Answer Quality and evaluated the system cost relative to the number of agents in the ensemble.\footnote{A notable limitation of these experiments was the inability to accurately measure system latency due to non-stationary external conditions, such as API rate limiting. Future work will involve simulating realistic conditions to accurately measure latency.}

\subsection{Preprocessing Methods}
We implemented two main types of preprocessing logic:
\begin{itemize}

\item Thresholding:
    \begin{itemize}
    \item Thresholding (Control): A naive truncation strategy where the first 8,000 tokens from the search results are taken.
    \item Vertical Thresholding: Selects top tokens from the top two verticals (until the prompt reaches 8,000 tokens).
    \item Aggressive Thresholding: Takes the first 6,000 tokens from the search response, allowing more space for the reasoning process.
    \end{itemize}
\item Re-ranking: Utilized a cross-encoder model based on `ms-marco-MiniLM-L-12-v2' from the Sentence Transformers library \cite{sentence-bert}.  The cross-encoder accepts the query and document as input and returns a relevance score between 0 and 1.  Documents are re-ranked based on relevance scores before thresholding.\footnote{Note: The raw entity JSON is not used directly as input. Search configurations define an LLM-friendly view of the entity, specifying which fields should be included. The design of this view generation model is outside the scope of this work.}
\end{itemize}

\subsection{Arbitration}
In our multi-agent RAG system, we employ an arbitration algorithm to select the optimal response from a set of candidate answers generated by various QA agents. The arbitration process is governed by a decision threshold $T\in(0,1)$ and an arbitration function $A$ that maps a set of results to a single final answer.  The algorithm is defined in Algorithm \ref{alg:arbitration}

\begin{algorithm}[ht]
\DontPrintSemicolon
\SetKwInput{KwInput}{Input}
\SetKwInput{KwOutput}{Output}

\KwInput{Set of candidate responses $C = \{ c_1, c_2, \dots, c_n \}$; threshold $T$}
\KwOutput{Final answer or a negative result}

\BlankLine
\textbf{Step 1: Candidate Response Collection}\;
Let $C$ be the set of candidate responses from $N$ QA agents.\;

\BlankLine
\textbf{Step 2: Affirmative Response Identification}\;
Define $C_{\text{affirmative}} \subseteq C$ as the subset of responses that include an attempted answer (i.e., non-null and substantive responses).\;

\BlankLine
\textbf{Step 3: Threshold Evaluation}\;
Compute the minimum number of affirmative responses required:\;
\[
k = \left\lfloor T \times |C| \right\rfloor
\]\;
\uIf{$|C_{\text{affirmative}}| \geq k$}{
    Proceed to arbitration.\;
}
\Else{
    Return a negative result indicating that no answer will be provided.\;
    \BlankLine
    \Return Negative Result\;
}

\BlankLine
\textbf{Step 4: Arbitration Function Application}\;
Apply the arbitration function $A$ to $C_{\text{affirmative}}$:\;
\[
\text{Final Answer} = A(C_{\text{affirmative}})
\]\;
\Return Final Answer\;

\caption{Candidate Response Collection and Arbitration}
\label{alg:arbitration}
\end{algorithm}

For our experiments, the arbitration function $A$ was implemented using a cross-encoder model, consistent with the post-retrieval preprocessing methods described earlier. The cross-encoder jointly encodes each candidate answer with the original query to compute a relevance score. It effectively re-ranks the affirmative responses based on their contextual relevance and quality, selecting the top-ranked answer as the final output.

\subsection{Algorithm Ablations}

To isolate the contributions of each component in our arbitration algorithm, we conducted a series of ablation experiments. First, we bypassed the arbitration function entirely by selecting a candidate answer at random from the set of affirmative responses. Next, we evaluated the effect of the threshold mechanism by randomly choosing a response from the affirmative set after threshold evaluation. Finally, we applied the cross-encoder-based arbitration function \(A\) over the complete set of candidate responses to assess its overall impact on the final output.

\section{Results}
In this section, we present the results of our experiments, focusing on both single-agent ablations and multi-agent configurations. Our goal is to evaluate how different preprocessing strategies and arbitration mechanisms affect the performance of the QA system, particularly in terms of precision, recall, hallucination rate, and incongruent response rate.

\subsection{Single-Agent Experiments}
The results of the single-agent experiments are summarized in the upper section of Table \ref{tab:experimental_results}.  These results encompass the three preprocessing strategies discussed in the Single Agent Experiments section

\textbf{Key Findings}
\begin{itemize}
\item Aggressive Thresholding emerged as the best-performing strategy, achieving the highest precision and recall, as well as the lowest incongruent response rate.
\item Vertical Thresholding performed comparably to aggressive thresholding, indicating that focusing on top-ranked verticals effectively preserves relevant context.
\item Agents utilizing Re-ranking with Cross-Encoder exhibited lower precision and recall. The cross-encoder tended to eliminate important context documents, leading to reduced precision despite a lower hallucination rate.
\end{itemize}

\begin{table*}[t]
\centering
\caption{Experimental Results for Single and Multi-Agent Experiments}
\label{tab:experimental_results}
\resizebox{\textwidth}{!}{%
\begin{tabular}{lccccc}
\toprule
\textbf{Experiment Version} & \textbf{Recall} & \textbf{Precision} & \textbf{F1} & \textbf{Hallucination Rate} & \textbf{Incongruent Response Rate} \\
\midrule
Thresholding (control) & 0.640 & 0.656 & 0.648 & 0.239 & 0.015 \\
Rerank with thresholding & 0.529 & 0.652 & 0.584 & \textbf{0.187} & 0.027 \\
Rerank with vertical thresholding & 0.571 & 0.619 & 0.594 & 0.249 & 0.017 \\
Aggressive Thresholding & \textbf{0.656} & \textbf{0.670} & \textbf{0.663} & 0.235 & \textbf{0.008} \\
Vertical Thresholding & \textbf{0.656} & 0.654 & 0.655 & 0.249 & 0.013 \\
\midrule
\texttt{vote\_with\_thresh, N=3} & 0.655 & 0.672 & 0.663 & 0.230 & \textbf{0.011} \\
\texttt{vote\_with\_thresh, N=5} & 0.\textbf{684} & \textbf{0.691} & \textbf{0.688} & \textbf{0.220} & 0.012 \\
\texttt{vote\_most\_relevant, N=3} & 0.666 & 0.674 & 0.670 & 0.229 & 0.014 \\
\texttt{vote\_most\_relevant, N=5} & \textbf{0.684} & 0.681 & 0.683 & 0.229 & 0.013 \\
\texttt{vote\_most\_relevant\_with\_thresh, N=3} & 0.652 & 0.672 & 0.661 & 0.228 & 0.013 \\
\texttt{vote\_most\_relevant\_with\_thresh, N=5} & \textbf{0.684} & \textbf{0.691} & \textbf{0.688} & \textbf{0.220} & 0.012 \\
\bottomrule
\end{tabular}
}
\end{table*}

\subsection{Multi-Agent Experiments}
\begin{figure}
\centerline{\includegraphics[width=0.85\linewidth]{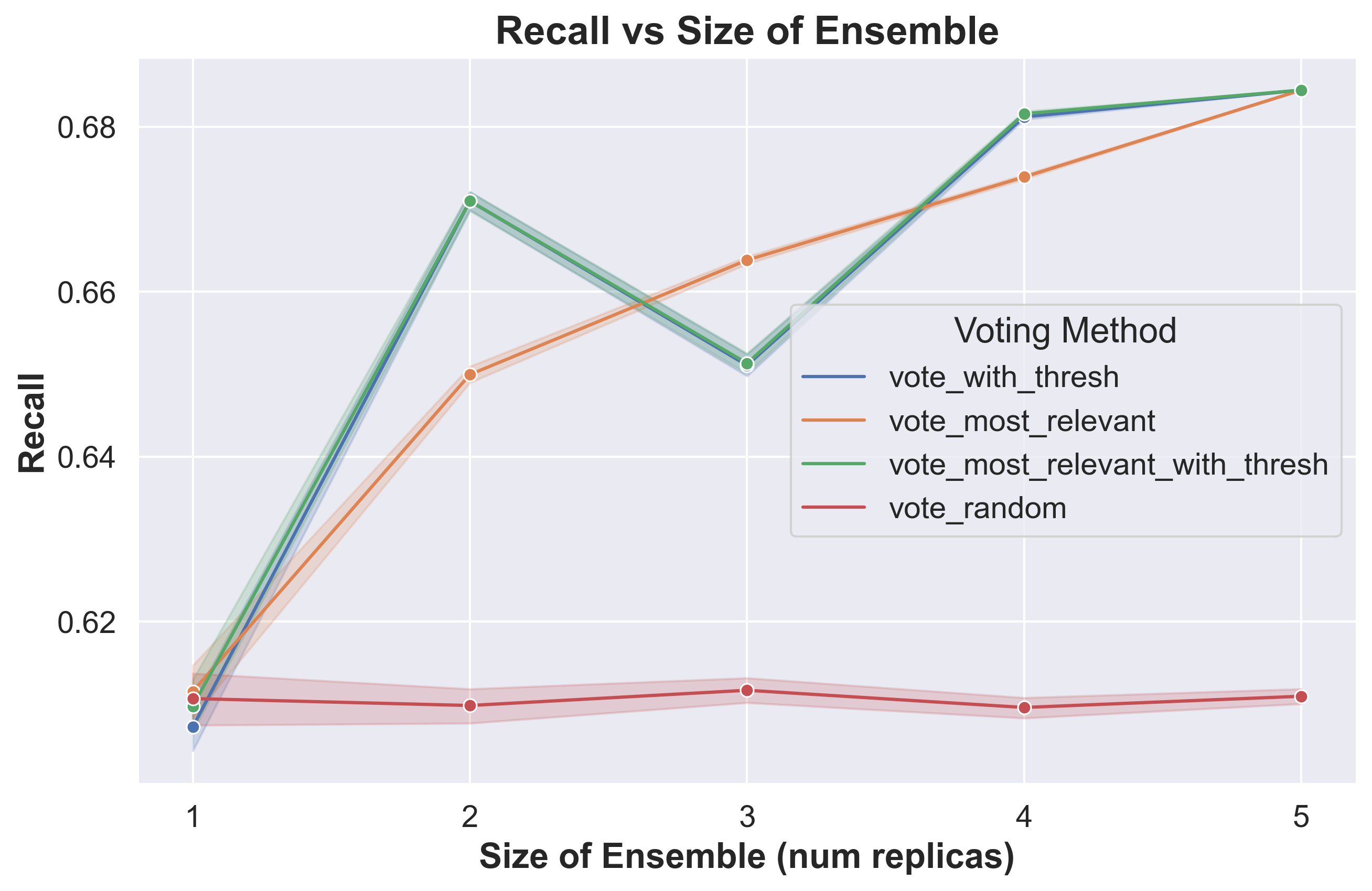}}
\caption{Recall vs. Ensemble Size}
\label{recall_chart}
\end{figure}

\begin{figure}
\centerline{\includegraphics[width=0.85\linewidth]{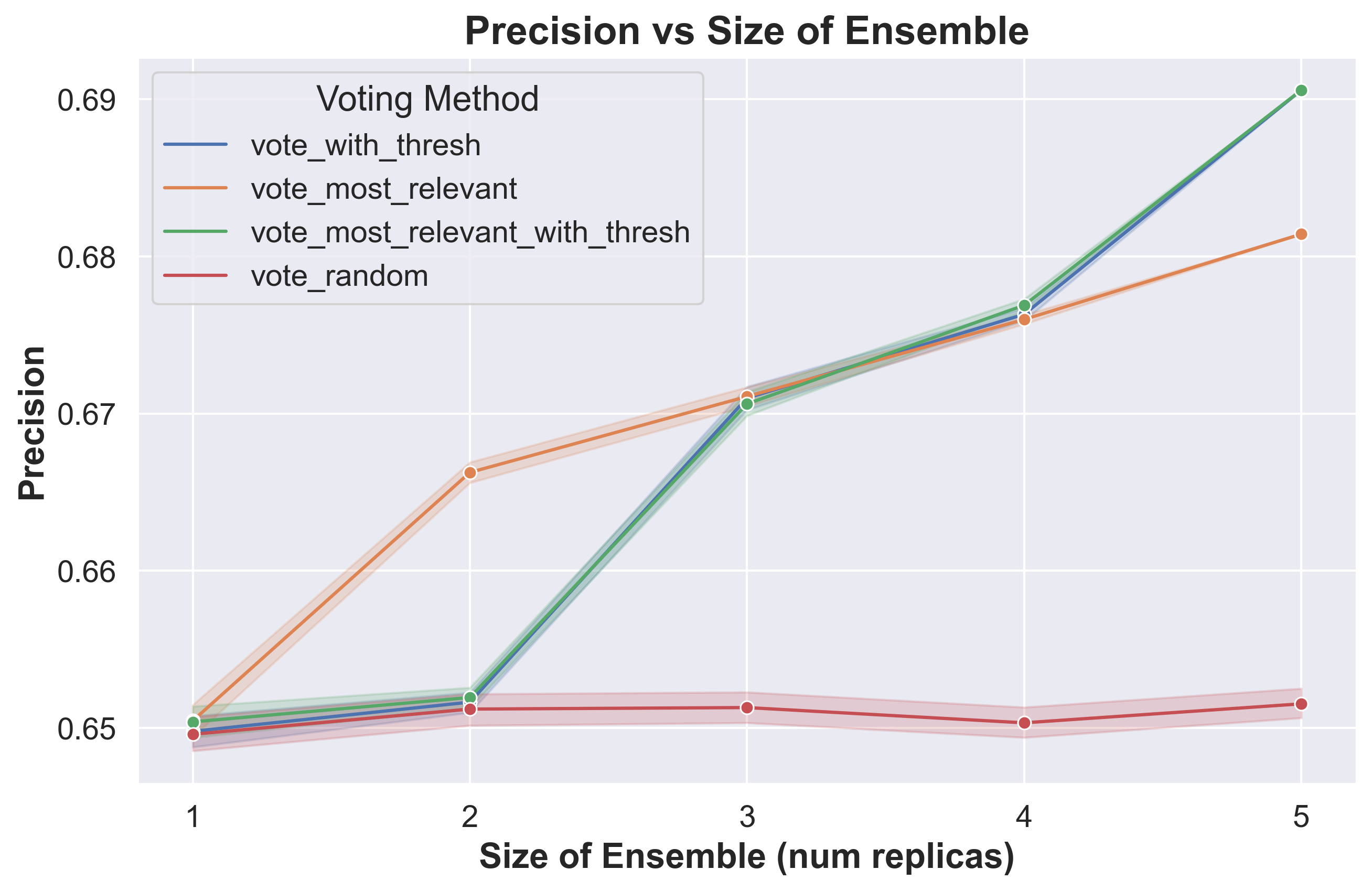}}
\caption{Precision vs. Ensemble Size}
\label{precision_chart}
\end{figure}

\begin{figure}
\centerline{\includegraphics[width=0.85\linewidth]{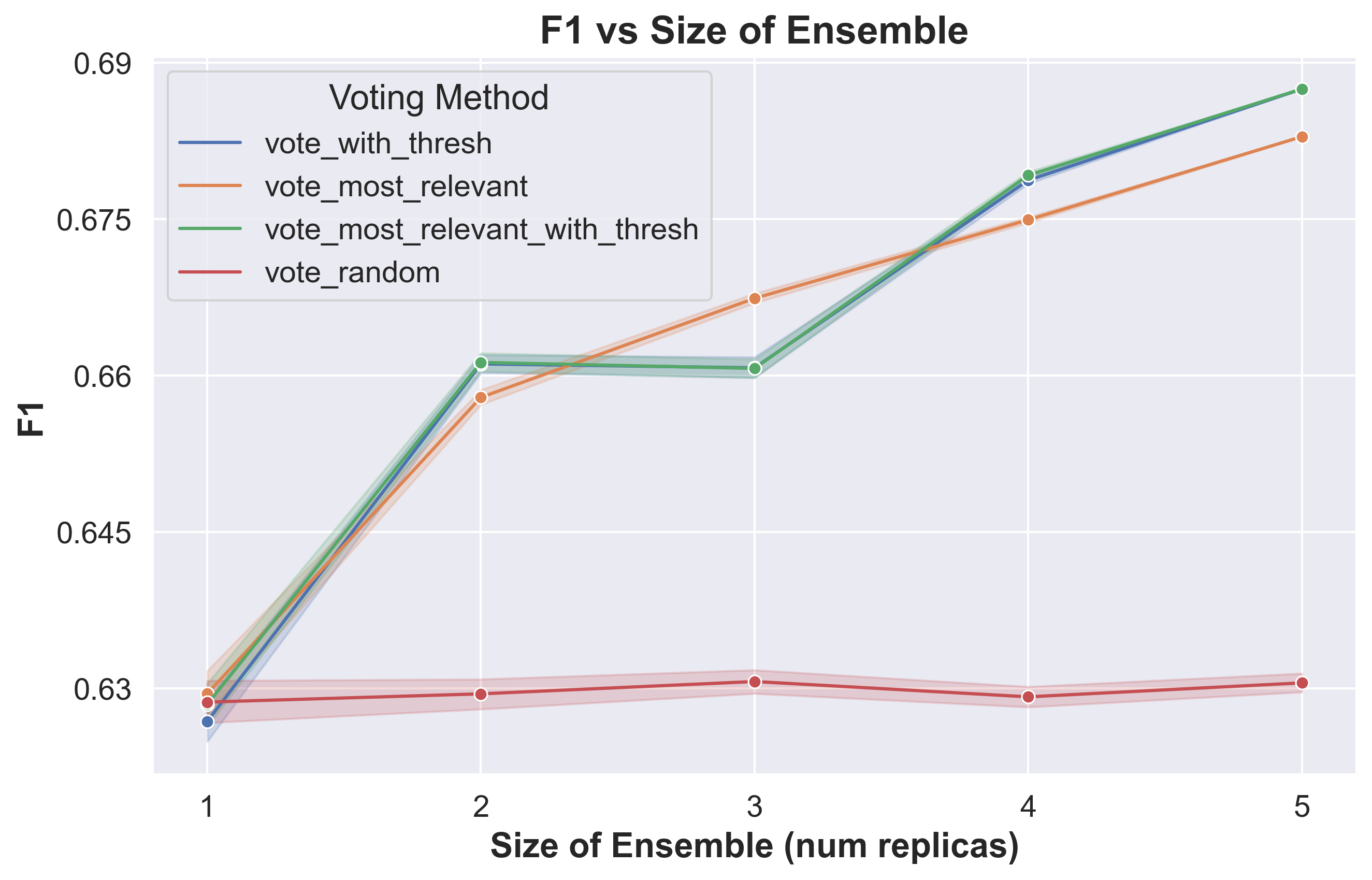}}
\caption{F1 vs. Ensemble Size}
\label{f1_chart}
\end{figure}

\begin{figure}
\centerline{\includegraphics[width=0.85\linewidth]{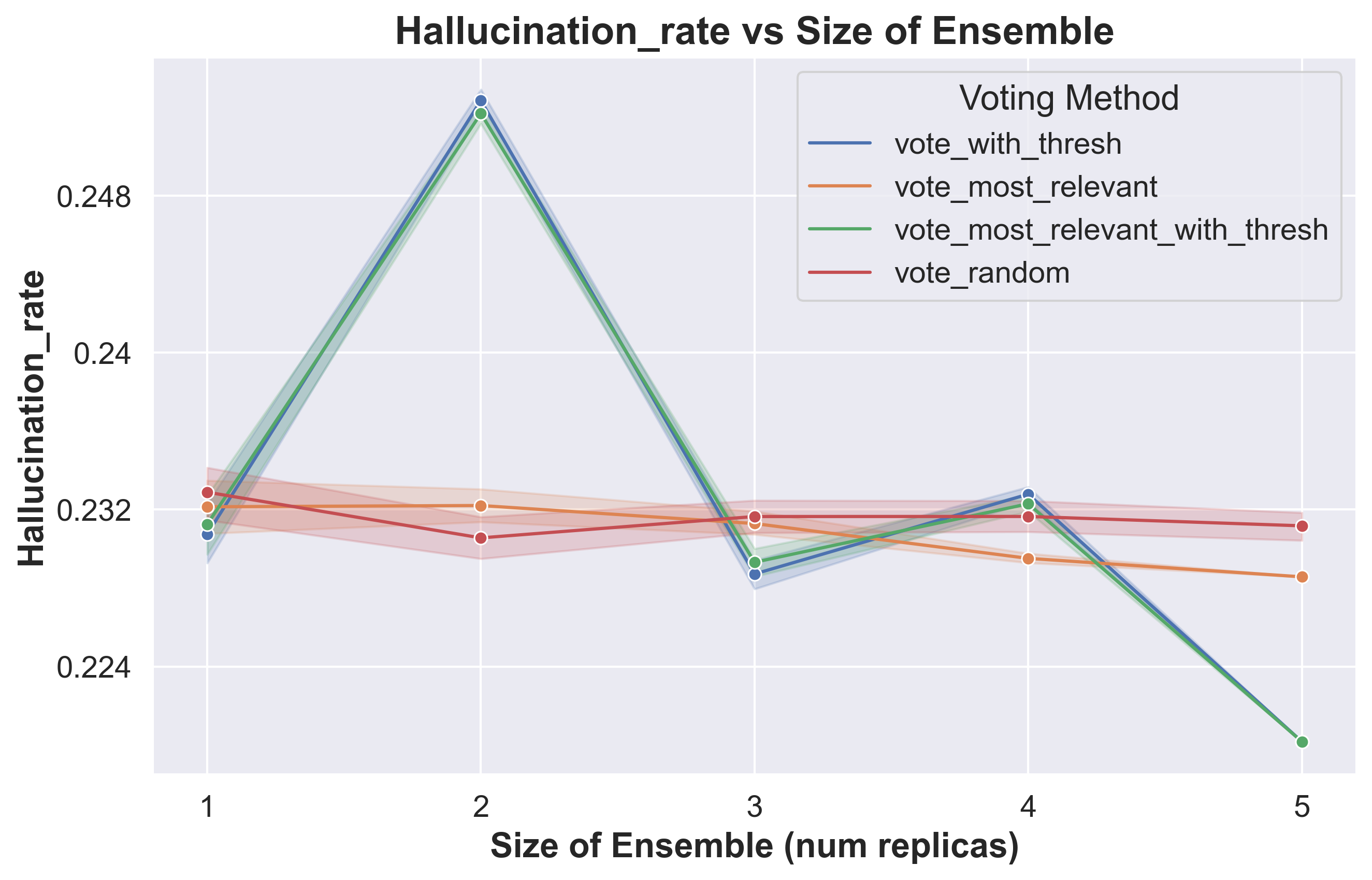}}
\caption{Hallucination Rate vs. Ensemble Size}
\label{hallucination_chart}
\end{figure}

\begin{figure}
\centerline{\includegraphics[width=0.85\linewidth]{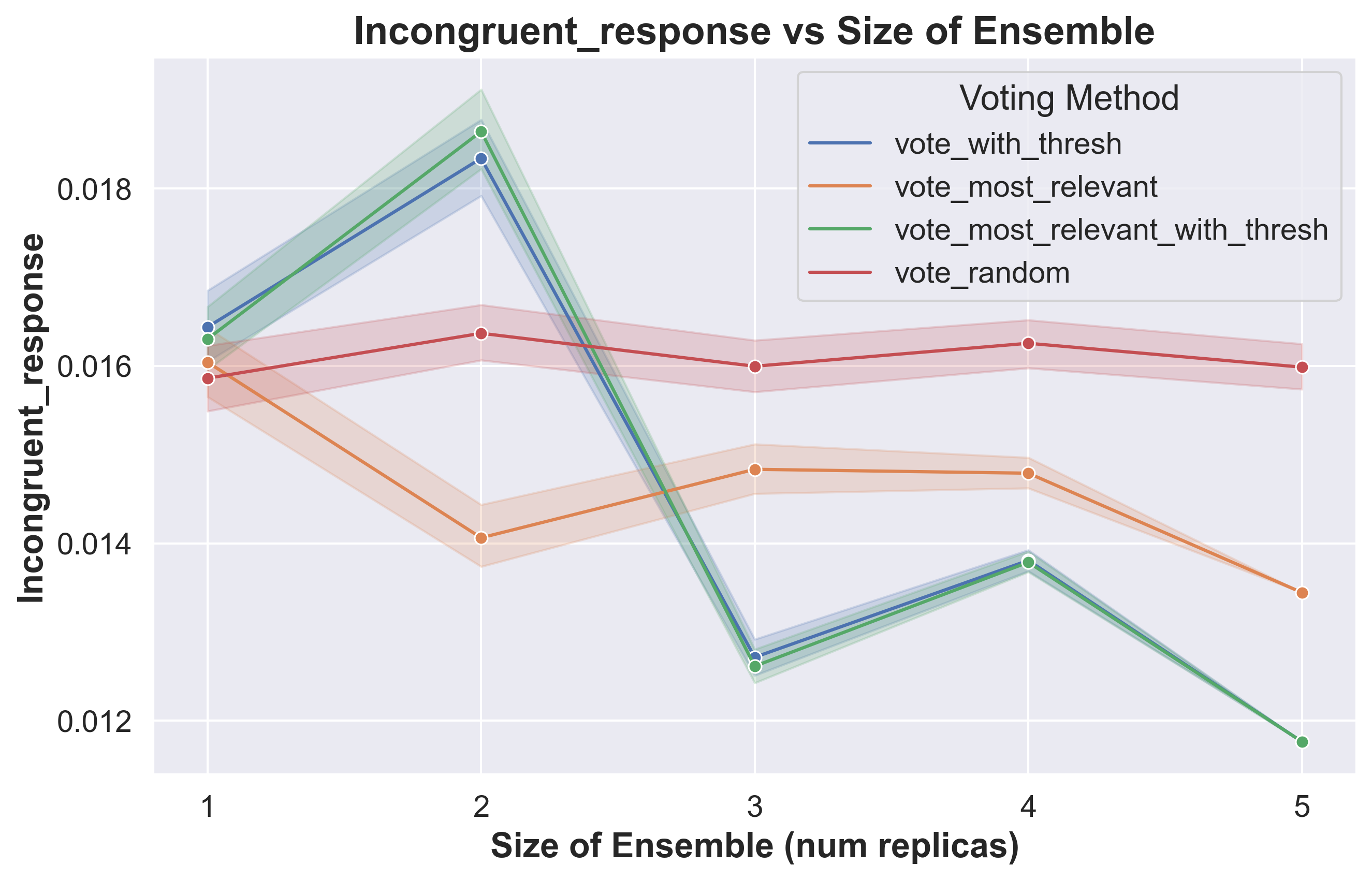}}
\caption{Incongruent Response Rate vs. Ensemble Size}
\label{irr_chart}
\end{figure}

The multi-agent experiments were designed to evaluate the impact of ensemble configurations on system performance. We varied the number of agents in the ensemble and observed the effects of different arbitration mechanisms, all of which used a threshold $T=0.5$, requiring at least half of the agents to provide an affirmative answer for a response to be returned.  The results can be found in the bottom section of Table \ref{tab:experimental_results}. Detailed charts are shown in Figures \ref{recall_chart} - \ref{irr_chart}.

\textbf{Key Findings}
\begin{itemize}
\item Increasing the number of agents consistently improved performance, enhancing precision and recall while reducing the hallucination and incongruent response rates. This highlights the benefit of aggregating responses across agents.
\item Balancing performance improvements with computational costs is crucial. Larger ensembles offer better performance metrics but come at the expense of increased resource usage and latency.
\item Simple voting mechanisms often performed on par with relevance-based arbitration. While the latter showed predictable improvements with more agents, the additional gains were minimal compared to the increased computational overhead.
\end{itemize}


\section{Discussion and Future Directions}

In this work, we introduce a dynamically reconfigurable multi-agent RAG system and present experimental results that highlight its ability to navigate multiple SLA dimensions—balancing answer quality and resource efficiency. Our results show that aggressive thresholding at the retrieval stage, by focusing on the most pertinent context, enables the system to generate more responses with improved precision. However, conventional re-ranking methods, while effective at prioritizing relevant documents, can sometimes filter out essential context and thereby impact the overall performance balance. Similarly, scaling the number of agents tends to boost metrics like precision and recall, but this comes with increased resource consumption. These interdependent trade-offs make the case for a system that can adapt in real time to diverse SLA requirements.

Looking ahead, several promising research directions are apparent. Exploring alternative arbitration mechanisms—such as weighted voting or additional machine learning-based approaches—may help optimize overall system performance, even if improvements in one dimension might be offset by challenges in another. Incorporating additional explicit QoS parameters—particularly latency and computational constraints—into our experiments would enable a more nuanced approach to SLA management. Coupling this with simulations of real-world conditions, such as network congestion and variable computational resources, and optimizing threshold parameters will be critical for comprehensively understanding their impact on latency and other performance dimensions.  We would also like to explore enhancements to our planning modules that enable temporal replication and dynamic strategy adjustments, which could further improve the system's overall adaptability.

Overall, our work demonstrates that a systems-oriented approach to SLA management in reconfigurable multi-agent RAG systems is not only feasible but also essential for deploying QA solutions that meet the multifaceted demands of real-world applications.


\bibliographystyle{ACM-Reference-Format}
\bibliography{references}

\end{document}